\documentclass[aps,prl,reprint,groupedaddress,longbibliography]{revtex4-2}
\usepackage{graphicx}
\usepackage{verbatim}
\usepackage{subfigure}
\usepackage[english]{babel}
\usepackage{siunitx}
\usepackage{pifont}
\usepackage{amsmath}
\usepackage{adjustbox}

\usepackage{mathtools} 
\usepackage{upgreek}
\usepackage{multirow}

\usepackage{xcolor}
\usepackage{graphicx}
\usepackage{parskip}
\usepackage[margin=0.75in]{geometry}
\begin{document}

\title{Imaging-assisted single-photon Doppler-free laser spectroscopy and the ionization energy of metastable triplet helium}
\author{Gloria Clausen$^1$, Simon Scheidegger$^1$, Josef A. Agner$^1$, Hansj\"urg Schmutz$^1$, Fr\'ed\'eric Merkt$^{1,2,3}$}
\affiliation{$^1$Department of Chemistry and Applied Biosciences, ETH Zurich, CH-8093 Zurich, Switzerland\\
	$^2$Quantum Center, ETH Zurich, CH-8093 Zurich, Switzerland\\
	$^3$Department of Physics, ETH Zurich, CH-8093 Zurich, Switzerland}

\date{\today}

\begin{abstract}
Skimmed supersonic beams provide intense, cold, collision-free samples of atoms and molecules are one of the most widely used tools in atomic and molecular laser spectroscopy. High-resolution optical spectra are typically recorded in a perpendicular arrangement of laser and supersonic beams to minimize Doppler broadening. Typical Doppler widths are nevertheless limited to tens of MHz by the residual transverse-velocity distribution in the gas-expansion cones. We present an imaging method to overcome this limitation which exploits the correlation between the positions of the atoms and molecules in the supersonic expansion and their transverse velocities - and thus their Doppler shifts. With the example of spectra of $(1{\rm s})(n{\rm p})\,^3{\rm P}_{0-2}\leftarrow (1{\rm s})(2{\rm s})\,^3{\rm S}_1$ transitions to high Rydberg states of metastable triplet He, we demonstrate the suppression of the residual Doppler broadening and a reduction of the full linewidths at half maximum to only about 1~MHz in the UV. Using a retro-reflection arrangement for the laser beam and a cross-correlation method, we determine Doppler-free spectra without any signal loss from the selection, by imaging, of atoms within ultranarrow transverse-velocity classes. As an illustration, we determine the ionization energy of triplet metastable He and confirm the significant discrepancy between recent experimental (Clausen {\it et al.}, Phys. Rev. Lett. {\bf 127} 093001 (2021)) and high-level theoretical (Patk{\'o}s {\it et al.}, Phys. Rev. A {\bf 103} 042809 (2021))  values of this quantity.
\end{abstract}
\maketitle

\begin{figure*}[t]
	\includegraphics[trim=0cm 0cm 0cm 0cm, clip=true, width=1.0\textwidth]{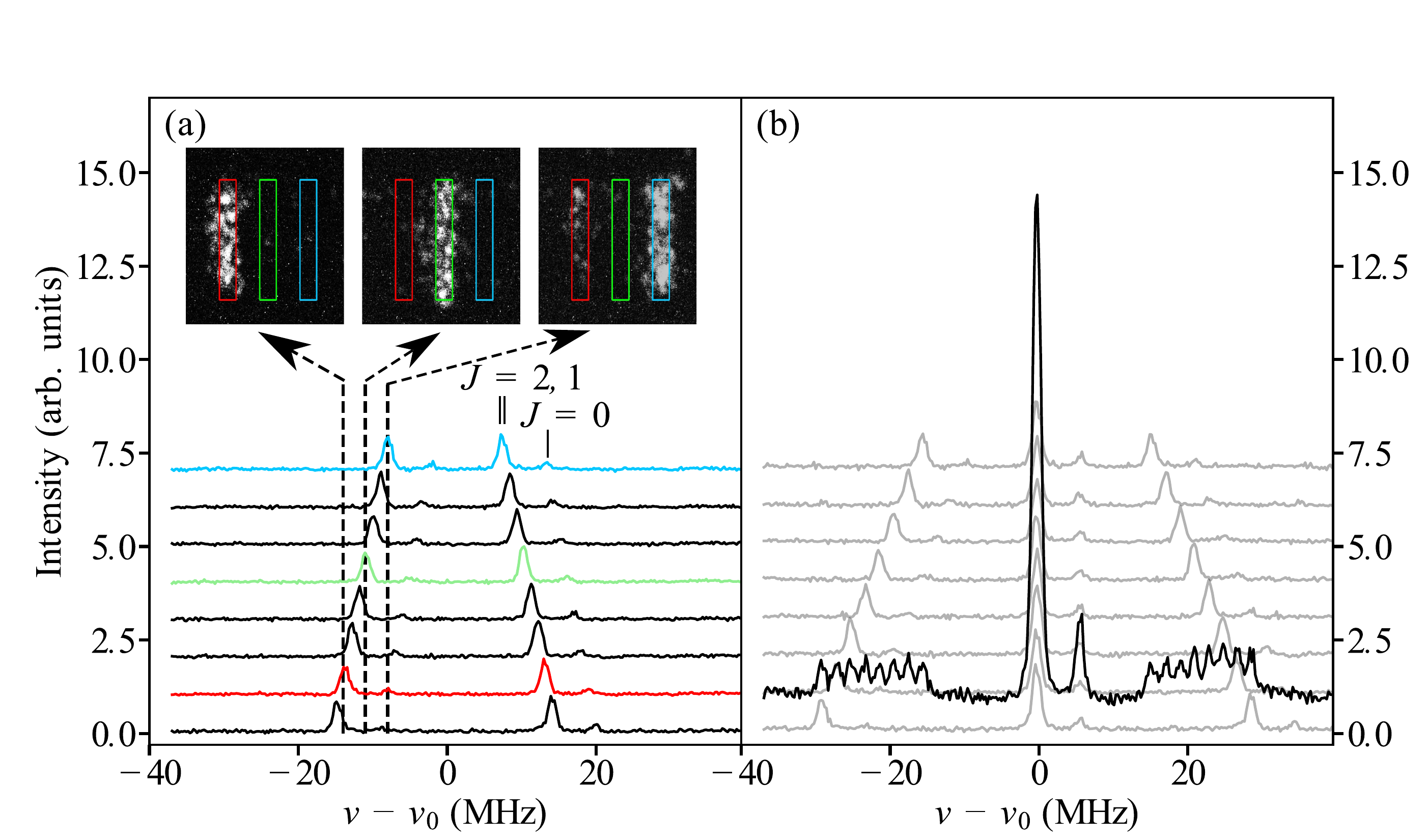}
	\caption{\label{fig:3} a) Sub-Doppler spectra of the $(1\mathrm s)(33\mathrm p)\,^3\mathrm P_{0-2}\leftarrow (1\mathrm s)(2\mathrm s)\,^3\mathrm S_1$ transitions in He recorded in adjacent areas of the imaging detector. The insets present the images recorded at the frequencies marked by dashed vertical lines for the three sub-Doppler spectra highlighted in red, green and blue. b) Gray: Cross-correlation spectra between the individual sub-Doppler spectra and synthetic spectra consisting of two Dirac delta distributions separated by the corresponding Doppler shift. Black: Sum of the cross-correlation spectra revealing the sharp Doppler-free transitions and their weak Doppler-broadened side bands.}
\end{figure*}

Atoms and molecules absorb and emit radiation at frequencies that appear shifted to higher or lower frequencies depending on whether they move towards or away from the observer. This effect is known as the Doppler effect and leads to a broadening of spectral lines in atomic and molecular spectra that reflects the velocity distribution of the sample of absorbing or emitting atoms and molecules. This line broadening represents a major challenge in high-resolution spectroscopy \cite{haensch83a}. Efforts towards eliminating Doppler broadening in atomic and molecular spectroscopy have a long history \cite{biraben19a} and can be classified in three categories. The first consists in reducing the spread of the velocity distribution in the direction of the observer by cooling the gas or forming a collimated gas beam. Remarkable examples in this category include full resolution of the rotational structure in the IR spectrum of C$_{60}$ after buffer-gas cooling \cite{changala19a} and precision-spectroscopic studies of free radicals in supersonic expansions \cite{kortyna18a}. In the second approach, referred to as sub-Doppler spectroscopy, spectra are recorded after selection of a subset of atoms or molecules belonging to a narrow velocity class out of the sample, for instance through double-resonance methods, resonant multiphoton excitation, and saturation spectroscopy \cite{demtroeder11a}, or through the detection of spatially resolved fluorescence \cite{lynds96a}. Recent examples include precision measurements of rovibrational transitions in CH$_4$ \cite{liu23a} and H$_2$O \cite{diouf22a}. In sub-Doppler spectroscopy, only a fraction of the sample contributes to the line intensities, which reduces the sensitivity.
In the third approach, the Doppler shifts are canceled through two-photon absorption from two counter-propagating laser beams of the same frequency. The measurements are thus Doppler-free. Precision measurements of the 1s-2s transition frequency in H are the best examples in this category \cite{parthey11a, matveev13a}. The great advantage of this approach beyond the elimination of the Doppler width is that all atoms or molecules in the sample contribute to the intensity of the Doppler-free lines, regardless of their velocity. Recent developments have combined two-photon Doppler-free spectroscopy with Ramsey-comb interferometry \cite{altmann18a} and broadband Fourier-transform dual-comb spectroscopy \cite{meek18a}. So far, no equivalent approach has been established for single-photon spectroscopy. The purpose of this letter is to fill this gap and introduce a method to record optical spectra with the best possible signal-to-noise ratio and spectral resolution. 
\\
\begin{figure*}[t]
\includegraphics[trim=0cm 0cm 0cm 0cm, clip=true, width=1.0\textwidth]{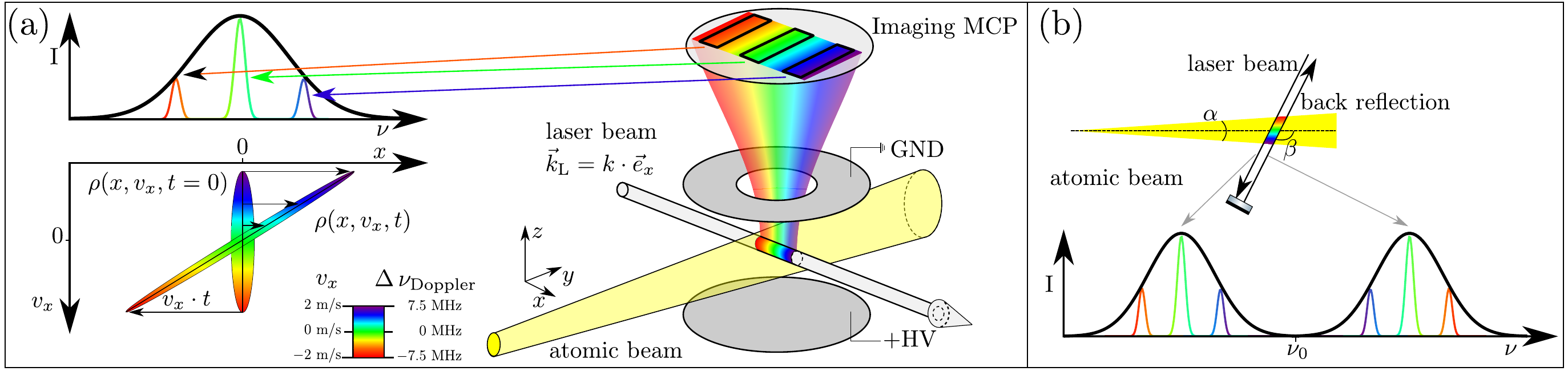}
\caption{\label{fig:1} Principle of imaging-assisted sub-Doppler spectroscopy in supersonic beams. a) Experimental geometry exploiting the correlation between the transverse velocities of the atoms and their positions along the laser beam, illustrated by the phase-space diagram. This correlation leads to Doppler-shift-dependent photoexcitation spots, symbolized by the rainbow-color scheme, which are imaged following perpendicular extraction of the ions generated by pulsed-field ionization. b) Determination of the Doppler-free transition frequency following inversion of the Doppler shifts upon retro-reflection of the laser beam. An intentional slight deviation from $90^\circ$ of the angle $\beta$ between the atomic and laser beams leads to pairs of lines with opposite Doppler shifts and a first-order-Doppler-free frequency $\nu_0$.}
\end{figure*}
\indent
The method exploits the correlation between the positions and velocities of atoms or molecules in an expanding gas sample and is schematically illustrated in Fig.~\ref{fig:1} with the example of the supersonic expansion used here as demonstration. 
The free expansion of the supersonic beam leads to a distribution of transverse
velocities along the laser propagation axis with position-specific
Doppler shifts of the atomic or molecular transition frequencies
indicated by the rainbow-color scheme. The corresponding phase-space evolution $\rho(x, v_x, t)$ is also represented schematically in Fig.~\ref{fig:1} (a). Pulsed-field ionization (PFI)
(or ionization with a second photon) of the upper level of the
transition leads to the generation of ions at positions determined by
the Doppler shifts. Extraction of the ions with a homogeneous-electric-field pulse towards a microchannel-plate (MCP) detector
connected to a charge-coupled-device (CCD) camera is used to spatially image \cite{chandler87a} the
generated ion cloud. The detection positions of the ions on the image are correlated
to the Doppler shifts. Monitoring the ionization signal in selected areas
of the image (indicated by rectangles) as a function of the laser frequency enables the recording of spectra with sub-Doppler
linewidths (colored lines in the top left spectrum). To determine the
Doppler-free transition frequency, the laser propagation axis is chosen
not to be exactly perpendicular to the gas-beam axis (intersection angle
$\beta \approx 89.5^\circ$) and is retro-reflected by a mirror located beyond the gas
beam, which inverts the Doppler shifts and leads to pairs of lines
(Fig.~\ref{fig:1} (b)). The Doppler-free transition frequency $\nu_0$ is the
average of the frequencies of the pair of corresponding sub-Doppler lines. A cross-correlation method is then used to combine all sub-Doppler spectra in a single Doppler-free spectrum of the full sample. In this letter, we demonstrate this method with a proof-of-principle measurement of the $(1\mathrm s)(n\mathrm{p})\,^3\mathrm{P}_J \leftarrow (1\mathrm s)(2\mathrm{s})\,^3\mathrm{S}_1$ transitions in $^4\mathrm{He}$. \\
\begin{figure}[b]
\includegraphics[trim=0cm 0cm 0cm 0cm, clip=true, width=1.0\columnwidth]{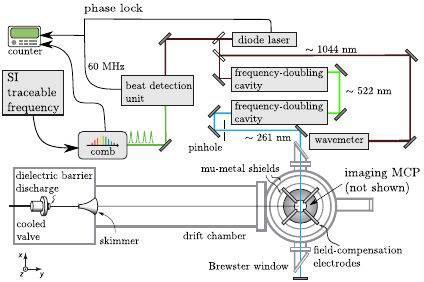}
\caption{\label{fig:2} Schematic diagram of the experimental setup. Top: Laser system including the single-mode frequency-quadrupled diode laser used to photoexcite He$^*$ to high Rydberg states and the frequency-comb-based SI-traceable frequency-calibration system. Bottom: Valve and discharge electrodes used to generate the supersonic expansion of He$^*$, drift chamber, and photoexcitation chamber, where the atomic and laser beams cross on the axis of a set of cylindrical electrodes used to field ionize the Rydberg states and extract the He$^+$ ions towards an imaging microchannel-plate (MCP) detector within a double-layer mu-metal magnetic shield and electrodes for three-dimensional electric-stray-field compensation.}
\end{figure}
\indent 
The experimental setup is displayed schematically in Fig. \ref{fig:2} and has been described earlier \cite{clausen21a}.
A supersonic beam of helium  atoms with velocities around $480\,\mathrm{m/s}$ is generated from a high-pressure reservoir emitting helium atoms into vacuum by a pulsed valve (repetition rate $8.33\,\mathrm{Hz}$) cooled to $12\,\mathrm{K}$. A dielectric-barrier discharge at the nozzle orifice excites $^4$He to the metastable $(1\mathrm{s)(2\mathrm{s}})\,^3\mathrm{S}_1$  state (He$^*$ hereafter). Transversely cold He$^*$ is selected with a 1-mm-diameter skimmer placed in a distance of $45\,\mathrm{cm}$ from the valve orifice. The skimmed atomic beam enters the interaction region $1.35\,\mathrm{m}$ downstream from the skimmer, where it is intersected by a single-mode continuous-wave UV laser beam ($\lambda \approx 260\,\mathrm{nm}$, waist diameter of $1.2\,\mathrm{mm}$, power $\approx 10\,\mathrm{mW}$) at near-right angles to excite $n$p Rydberg states. The transitions are detected by PFI of the Rydberg states and extraction of the $^4\mathrm{He}^+$ ions toward an MCP imaging detector in the direction perpendicular to the atomic and laser beams ($z$ axis). The photoexcitation region is magneticlly shielded using two concentric mumetal shields. Stray electric fields are compensated to below $1\,\mathrm{mV/cm}$ in all three spatial directions using a stack of plane electrodes in the $z$ direction and four pin electrodes in the $x$ and $y$ directions \cite{clausen21a}.  
\\ \indent 
The phosphorescence of the imaging MCP is detected with a CCD camera. The resulting image is divided into adjacent detection areas along the direction of the laser beam (see Fig. \ref{fig:1} (a)). Spectra of the $(1\mathrm s)(n\mathrm{p})\,^3\mathrm{P}_J \leftarrow (1\mathrm s)(2\mathrm{s})\,^3\mathrm{S}_1$ transition are recorded by scanning the laser frequency, collecting the ions impinging on each of the detection areas separately, and monitoring the corresponding integrated signals.   
\\ \indent
The UV radiation is generated from the output of a tapered-amplifier diode laser at $1040\,\mathrm{nm}$, which is frequency quadrupled using two doubling cavities  equipped with nonlinear frequency-upconversion crystals. A 1\% reflection of the laser beam is used to reduce the laser linewidth to $\ll 80\,\mathrm{kHz}$ by a phase lock to an optical ultra-low-noise frequency comb (\textit{FC1500-250-ULN, Menlo
Systems}). The 1040~nm laser is frequency shifted with an acousto-optic modulator (AOM) operated in a double-pass configuration, allowing one to scan the fundamental laser frequency with respect to the frequency comb. The comb itself is short-term stabilized by a 1572-nm laser locked to a high-finesse cavity. Long-term drifts are detected by referencing the frequency comb to an ultrastable and SI-traceable frequency disseminated by the stabilized fiber network described in Ref. \onlinecite{husmann21a}. The UV laser frequency is scanned in discrete steps $\Delta \nu$ (typically $\Delta \nu=240\,\mathrm{kHz}$) and, at each step, the data are averaged over 75 experimental cycles.
To cancel the first-order Doppler shift, the laser beam is retroreflected and a distant pinhole is used to ensure a reflection angle within $180.00 \pm 0.03^\circ$. 
\\ \indent
Fig. \ref{fig:3} (a) illustrates the measurement procedure with spectra of the $(1\mathrm s)(33\mathrm p)\,^3\mathrm P_{0-2}\leftarrow (1\mathrm s)(2\mathrm s)\,^3\mathrm S_1$ transitions obtained from eight adjacent detector regions. For clarity, the different spectra were shifted along the $y$ axis according to the spatial ordering of the detection areas. The spin-orbit interaction splits the $33\mathrm{p}$ triplet state into three components with $J=0-2$, two of which ($J=1,2$) are almost degenerate, as assigned in the top spectrum of Fig. \ref{fig:3} (a) \cite{drake99a}. Each spectrum consists of two Doppler components having opposite first-order Doppler shifts. Because each selected area of the images corresponds to a specific transverse-velocity class, the splitting between the Doppler components evolves from one spectrum to the next. \\
\indent 
The three images presented as insets in Fig. \ref{fig:3} (a) illustrate the efficacy of the selection of narrow velocity classes. They were recorded at the frequencies corresponding to the vertical dashed lines. At the lowest of these frequencies, the He$^+$ ions are detected on the left of the image in the area used to record the spectrum drawn in red. At the middle frequency, the He$^+$ ions are detected near the image center in the area used to record the spectrum drawn in green. In both cases, the detected He$^+$ ions result from PFI of Rydberg atoms in the $J=1,2$ fine-structure components. At the highest frequency, He$^+$ ions are detected both on the left and the right of the image, corresponding to the weak transition to the $J=0$ component of the spectrum drawn in red and the stronger transitions to the $J=1,2$ components of the spectrum drawn in blue, respectively. 
\\ \indent
To obtain a Doppler-free spectrum that combines the information contained in the spectra obtained from all selected detector areas, we use a cross-correlation method. 
All spectra in Fig. \ref{fig:3} (a) consist of $N$ data points $f(\nu_i)=f(i)$ with index $i=0,1,...,N-1$ at frequencies $\nu_i=\nu_\mathrm{start}+i \Delta \nu$. We evaluate the circular cross correlation $R_{fg}$ between each spectrum and a periodically-defined test function $g(\nu_i)=g(i)$ consisting of two Dirac delta distributions separated by a frequency interval $\nu_d=d \Delta \nu$ according to \begin{equation}
R_{fg}(k)=\sum_{i=0}^{N-1} f(i) g(i+k)_\mathrm{mod\,N}.
\end{equation}
If $\nu_d$ does not match the Doppler splitting, each cross-correlation spectrum comprises four components of equal intensity per spectral line, because the cross-correlation leads to two lines separated by $\nu_d$ for each Doppler component. If $\nu_d$ exactly matches the Doppler splitting, the cross-correlation spectrum consists of two side bands separated by $2\nu_d$ and a central peak with double intensity at the Doppler-free position $\nu_0$ (see spectra drawn in gray in Fig. \ref{fig:3} (b)). We calculate cross-correlation spectra for gradually increasing $d$ values and determine, for all spectra in Fig. \ref{fig:3} (a), the Doppler splittings numerically as the $\nu_d$ values corresponding to the cross-correlation spectra with maximal peak intensity.
These cross-correlation spectra are depicted in gray in Fig. \ref{fig:3} (b). Their sum, drawn in black, represents a Doppler-free spectrum to which \textbf{all detected ions} contribute, regardless of the Doppler shift. The sum spectrum also contains Doppler-broadened side bands. The central Doppler-free lines have full widths at half maximum of only $\approx 1.2\,\mathrm{MHz}$. The large gain in signal-to-noise ratio resulting from the cross-correlation analysis is best recognized if one considers that the weak transition to the $(1\mathrm s)(33\mathrm{p})\,^3\mathrm{P}_0$ component is barely visible in the individual spectra of Fig. \ref{fig:3} (a). \\ \indent
To extract the centroid transition frequencies of the $(1\mathrm s)(n\mathrm p)\,^3\mathrm P_{0-2}\leftarrow (1\mathrm s)(2\mathrm s)\,^3\mathrm S_1$ transitions, we fit a sum of three Gaussian profiles, accounting for the three fine-structure components of the spectrum, with relative intensities of $5:3:1$ for the $J=2,\, J=1$ and $J=0$ components, respectively. The fit parameters describing the three-peak structure comprise the centroid frequency, the frequency differences between the peaks of the $J=0$ and $J=1$ and the $J=1$ and $J=2$ components, the maximum intensity, a linewidth determining the full width at half maximum (FWHM), and an overall background offset. \\  \indent
We measured transitions to two $n$p Rydberg states with principal quantum numbers $n=33$ and $n=40$ and repeated the measurements 20 times for the $(1{\rm s})(33{\rm p})\,^3{\rm P}_{0-2}\leftarrow (1{\rm s})(2{\rm s})\,^3{\rm S}_1$ transition and 10 times for the $(1{\rm s})(40{\rm p})\,^3{\rm P}_{0-2}\leftarrow (1{\rm s})(2{\rm s})\,^3{\rm S}_1$ transition. Before each new measurement, we realigned the laser beam and its back reflection. In this way, we transformed a systematic error from a possible misalignment angle between the laser beam and its back reflection into a statistical error, as explained in Ref. \onlinecite{clausen21a}.\\ \indent
The centroid transition frequency for a given transition is obtained as the weighted average of all measured spectra, and its statistical uncertainty is obtained from the weighted standard deviation of the averaged centroid frequency.
The statistical uncertainties ($112$ and $174\,\mathrm{kHz}$ for the transitions to $n=33$p and $n=40$p, respectively) are dominated by the residual first-order Doppler shift resulting from the uncertainty in the back-reflection angle. \\ \begin{table*}[t]
\caption{\label{Tab:1} Centroid transition frequencies for the $(1{\rm s})(33{\rm p})\,^3{\rm P}_{0-2}\leftarrow (1{\rm s})(2{\rm s})\,^3{\rm S}_1$ and $(1{\rm s})(40{\rm p})\,^3{\rm P}_{0-2}\leftarrow (1{\rm s})(2{\rm s})\,^3{\rm S}_1$ transitions, corrected for the photon-recoil shift and the second-order Doppler shift (second column), and centroid binding energies of the $(1{\rm s})(33{\rm p})\,^3{\rm P}$ and $(1{\rm s})(40{\rm p})\,^3{\rm P}$ Rydberg states, calculated using Eqs. (\ref{eq:2}) and (\ref{eq:3}) with the quantum-defect parameters taken from Ref.~\onlinecite{drake99a} (third column). Last column: Comparison of experimental and theoretical values of the ionization energies of the $(1{\rm s})(2{\rm s})\,^3{\rm S}_1$ state.}
\begin{ruledtabular}
\begin{tabular}{r lll }
&$\nu_\mathrm{trans}$ (centroid) / MHz& $E_\mathrm{binding}/h$ (centroid) / MHz  &$E_\mathrm{I}/h$ (2 $^3\mathrm{S}_1$) / (MHz) \\

$n$=33 &1149 809 632.350(112)& 3033 110.47333(11)& 1152 842 742.823(113)\\
$n$=40 &1150 779 829.994(174)&2062 912.80159(5)& 1152 842 742.796(174)\\
\citet{patkos21a} &n.a.& n.a.&1152 842 742.231(52) \\
\citet{clausen21a} &n.a. & n.a.&1152 842 742.640(32) \\
\end{tabular}
\end{ruledtabular}
\end{table*} \indent
We considered additional sources for systematic effects, as discussed in Ref. \onlinecite{clausen21a}. The photon-recoil shift and the second-order Doppler shift, $734\,\mathrm{kHz}$ and $-1.5\,\mathrm{kHz}$, respectively, are subtracted from the measured transition frequencies to yield the corrected transition frequencies listed in Tab. \ref{Tab:1}. Systematic uncertainties resulting from the pressure shift, DC-Stark shifts from the $\le 1$~mV/cm residual stray field, and the frequency calibration are negligible at our current precision. 
No change of the centroid transition frequencies could be detected when the excitation-laser power was reduced by a factor of 5, implying that AC-Stark shifts are also negligible. Momentum transfer by absorption of a photon from the incoming and reflected beam changes the transverse velocity by $\Delta v=\frac{\pm h \nu}{m}$, which affects the detection positions slightly \cite{wen23b}. Under our experimental conditions, this effect results in a shift of $\leq 1\,\mathrm{kHz}$ and can be neglected.\\ \indent
The ionization energy $E_{\mathrm I}(2\,^3\mathrm S_1)$ of the $ (1{\rm s})(2{\rm s})\ ^3{\rm S}_1$ state of helium is determined by combining the experimental transition frequencies $\nu_n=E_n/h$ with the binding energy of the 33p and 40p Rydberg states given by the Rydberg-Ritz formula \cite{ritz03a} \begin{equation} \label{eq:2}
E_{\mathrm I}(2\,^3\mathrm S_1)/h= E_n/h + \frac{R_\mathrm{He}\cdot c}{(n^{*})^{2}},
\end{equation}
where $R_\mathrm{He}=109\,722.275\,548\,36(21)\;\mathrm{cm}^{-1}$ is the mass-corrected Rydberg constant calculated according to Ref. \onlinecite{codata2018a}, and $n^*=n-\delta(n)$. The $n$-dependent quantum defect $\delta(n)$ is calculated iteratively using \cite{drake91a} 
\begin{equation} \label{eq:3} \begin{split}
&\delta(n)=  \delta_0+\frac{\delta_2}{(n-\delta(n))^2}+\\ &\frac{\delta_4}{(n-\delta(n))^4}+\frac{\delta_6}{(n-\delta(n))^6}+\frac{\delta_8}{(n-\delta(n))^8},\end{split}
\end{equation} 
with the quantum-defect parameters reported by Drake \cite{drake99a}. 
We independently verified the reliability of the quantum defects reported in Ref. \onlinecite{drake99a} for the $J=0$ fine-structure component, which is well resolved in our spectra.
\\ \indent
Our new results for the ionization energy of the $2\,^3\mathrm S_1$ state of $^4$He$^*$ (see Table~\ref{Tab:1}) agree within the uncertainties with our previous experimental result \cite{clausen21a}, 
which was obtained by combining the ionization energy of the $2\,^1\mathrm S_0$ state with the $2\,^1\mathrm S_0 \leftarrow 2\,^3\mathrm S_1 $ interval measured by Rengelink {\it et al.}~\cite{rengelink18a}, but differ from the theoretical value by $592(124)\,\mathrm{kHz}$ ($n=33)$ and $538(182)\,\mathrm{kHz}$ ($n=40$), confirming the discrepancy between theory and experiment for this fundamental quantity noted earlier \cite{clausen21a,patkos21a,yerokhin22a, yerokhin23a}. The calculations reported in Ref. \onlinecite{patkos21a} were the first to include the full $\alpha^7m$ Lamb shift in a two-electron system and this discrepancy currently prevents the determination of the $^{4}$He$^{2+}$-particle radius from He spectroscopy. 
\\ \indent
The method presented in this letter features all advantages of Doppler-free spectroscopy by combining three elements: (i) Imaging enables the subdivision of the Doppler-broadened line profiles into a multitude of narrow sub-Doppler line profiles by selecting atoms or molecules belonging to different velocity classes; (ii) the retroreflection arrangement of the laser beam and the slight deviation from 90$^\circ$ of the laser and supersonic beams permit the determination of Doppler-free transition frequencies; and (iii) the cross-correlation analysis combines the sub-Doppler spectra into a Doppler-free spectrum to which all atoms or molecules in the sample contribute. 
\\ \indent The method is not limited to the detection of Rydberg states by PFI. Without any change, it can be applied to resonance-enhanced two-photon ionization spectroscopy [(1+1) REMPI] in molecular beams, which is one of the most widely used techniques in molecular spectroscopy \cite{ashfold17a}. It can also be applied to laser-induced-fluorescence (LIF) spectroscopy \cite{zare11a}, which is another very broadly used technique in molecular spectroscopy, if the spatial imaging of the ions is replaced by spatial imaging of the fluorescence.
In the experiments presented here, we exploited spatial imaging to obtain sub-Doppler resolution. The spectral resolution could still be improved significantly by optimizing the ion-imaging optics and the gas-beam geometry. In this context, it is worth noting that the method might also find applications in a velocity-map-imaging \cite{eppink97a} mode of dissociation products.

\begin{acknowledgments}
This work was supported by the Swiss National Science Foundation through the Sinergia-program (Grant No. CRSII5-183579) and a single-investigator grant (Grant No. 200020B-200478).
\end{acknowledgments}

%

\begin{thebibliography}{27}%
\makeatletter
\providecommand \@ifxundefined [1]{%
 \@ifx{#1\undefined}
}%
\providecommand \@ifnum [1]{%
 \ifnum #1\expandafter \@firstoftwo
 \else \expandafter \@secondoftwo
 \fi
}%
\providecommand \@ifx [1]{%
 \ifx #1\expandafter \@firstoftwo
 \else \expandafter \@secondoftwo
 \fi
}%
\providecommand \natexlab [1]{#1}%
\providecommand \enquote  [1]{``#1''}%
\providecommand \bibnamefont  [1]{#1}%
\providecommand \bibfnamefont [1]{#1}%
\providecommand \citenamefont [1]{#1}%
\providecommand \href@noop [0]{\@secondoftwo}%
\providecommand \href [0]{\begingroup \@sanitize@url \@href}%
\providecommand \@href[1]{\@@startlink{#1}\@@href}%
\providecommand \@@href[1]{\endgroup#1\@@endlink}%
\providecommand \@sanitize@url [0]{\catcode `\\12\catcode `\$12\catcode
  `\&12\catcode `\#12\catcode `\^12\catcode `\_12\catcode `\%12\relax}%
\providecommand \@@startlink[1]{}%
\providecommand \@@endlink[0]{}%
\providecommand \url  [0]{\begingroup\@sanitize@url \@url }%
\providecommand \@url [1]{\endgroup\@href {#1}{\urlprefix }}%
\providecommand \urlprefix  [0]{URL }%
\providecommand \Eprint [0]{\href }%
\providecommand \doibase [0]{https://doi.org/}%
\providecommand \selectlanguage [0]{\@gobble}%
\providecommand \bibinfo  [0]{\@secondoftwo}%
\providecommand \bibfield  [0]{\@secondoftwo}%
\providecommand \translation [1]{[#1]}%
\providecommand \BibitemOpen [0]{}%
\providecommand \bibitemStop [0]{}%
\providecommand \bibitemNoStop [0]{.\EOS\space}%
\providecommand \EOS [0]{\spacefactor3000\relax}%
\providecommand \BibitemShut  [1]{\csname bibitem#1\endcsname}%
\let\auto@bib@innerbib\@empty
\bibitem [{\citenamefont {H\"ansch}(1983)}]{haensch83a}%
  \BibitemOpen
  \bibfield  {author} {\bibinfo {author} {\bibfnamefont {T.~W.}\ \bibnamefont
  {H\"ansch}},\ }\bibfield  {title} {\bibinfo {title} {Sub-{Doppler}
  {Spectroscopy}},\ }in\ \href@noop {} {\emph {\bibinfo {booktitle} {Atomic
  {Physics}}}},\ \bibinfo {series} {Proceedings of the Eighth International
  Conference on Atomic Physics, held August 2-6, 1982, at Chalmers University
  of Technology, Goteborg, Sweden}, Vol.~\bibinfo {volume} {8},\ \bibinfo
  {editor} {edited by\ \bibinfo {editor} {\bibfnamefont {I.}~\bibnamefont
  {Lindgren}}, \bibinfo {editor} {\bibfnamefont {A.}~\bibnamefont {Rosén}},\
  and\ \bibinfo {editor} {\bibfnamefont {S.}~\bibnamefont {Svanberg}}}\
  (\bibinfo  {publisher} {Springer US},\ \bibinfo {address} {Boston, MA},\
  \bibinfo {year} {1983})\ pp.\ \bibinfo {pages} {55--70}\BibitemShut {NoStop}%
\bibitem [{\citenamefont {Biraben}(2019)}]{biraben19a}%
  \BibitemOpen
  \bibfield  {author} {\bibinfo {author} {\bibfnamefont {F.}~\bibnamefont
  {Biraben}},\ }\bibfield  {title} {\bibinfo {title} {The first decades of
  {Doppler}-free two-photon spectroscopy},\ }\href
  {https://doi.org/https://doi.org/10.1016/j.crhy.2019.04.003} {\bibfield
  {journal} {\bibinfo  {journal} {Comptes Rendus Physique}\ }\textbf {\bibinfo
  {volume} {20}},\ \bibinfo {pages} {671} (\bibinfo {year} {2019})}\BibitemShut
  {NoStop}%
\bibitem [{\citenamefont {Changala}\ \emph {et~al.}(2019)\citenamefont
  {Changala}, \citenamefont {Weichman}, \citenamefont {Lee}, \citenamefont
  {Fermann},\ and\ \citenamefont {Ye}}]{changala19a}%
  \BibitemOpen
  \bibfield  {author} {\bibinfo {author} {\bibfnamefont {P.~B.}\ \bibnamefont
  {Changala}}, \bibinfo {author} {\bibfnamefont {M.~L.}\ \bibnamefont
  {Weichman}}, \bibinfo {author} {\bibfnamefont {K.~F.}\ \bibnamefont {Lee}},
  \bibinfo {author} {\bibfnamefont {M.~E.}\ \bibnamefont {Fermann}},\ and\
  \bibinfo {author} {\bibfnamefont {J.}~\bibnamefont {Ye}},\ }\bibfield
  {title} {\bibinfo {title} {Rovibrational quantum state resolution of the
  {C}$_{60}$ fullerene},\ }\href {https://doi.org/10.1126/science.aav2616}
  {\bibfield  {journal} {\bibinfo  {journal} {Science}\ }\textbf {\bibinfo
  {volume} {363}},\ \bibinfo {pages} {49} (\bibinfo {year} {2019})}\BibitemShut
  {NoStop}%
\bibitem [{\citenamefont {Kortyna}\ \emph {et~al.}(2018)\citenamefont
  {Kortyna}, \citenamefont {Lesko},\ and\ \citenamefont
  {Nesbitt}}]{kortyna18a}%
  \BibitemOpen
  \bibfield  {author} {\bibinfo {author} {\bibfnamefont {A.}~\bibnamefont
  {Kortyna}}, \bibinfo {author} {\bibfnamefont {D.~M.~B.}\ \bibnamefont
  {Lesko}},\ and\ \bibinfo {author} {\bibfnamefont {D.~J.}\ \bibnamefont
  {Nesbitt}},\ }\bibfield  {title} {\bibinfo {title} {High-resolution
  sub-{Doppler} infrared spectroscopy of atmospherically relevant {Criegee}
  precursor {CH$_2$I} radicals: {CH$_2$} stretch vibrations and
  “charge-sloshing” dynamics},\ }\href {https://doi.org/10.1063/1.5028287}
  {\bibfield  {journal} {\bibinfo  {journal} {The Journal of Chemical Physics}\
  }\textbf {\bibinfo {volume} {148}},\ \bibinfo {pages} {174308} (\bibinfo
  {year} {2018})}\BibitemShut {NoStop}%
\bibitem [{\citenamefont {Demtr{\"o}der}(2011)}]{demtroeder11a}%
  \BibitemOpen
  \bibfield  {author} {\bibinfo {author} {\bibfnamefont {W.}~\bibnamefont
  {Demtr{\"o}der}},\ }\bibfield  {title} {\bibinfo {title} {Doppler-free laser
  spectroscopy},\ }in\ \href
  {https://doi.org/https://doi.org/10.1002/9780470749593.hrs082} {\emph
  {\bibinfo {booktitle} {Handbook of High‐resolution Spectroscopy}}},\
  Vol.~\bibinfo {volume} {3},\ \bibinfo {editor} {edited by\ \bibinfo {editor}
  {\bibfnamefont {M.}~\bibnamefont {Quack}}\ and\ \bibinfo {editor}
  {\bibfnamefont {F.}~\bibnamefont {Merkt}}}\ (\bibinfo  {publisher} {John
  Wiley \& Sons},\ \bibinfo {address} {Chichester},\ \bibinfo {year} {2011})\
  pp.\ \bibinfo {pages} {1759--1780}\BibitemShut {NoStop}%
\bibitem [{\citenamefont {Lynds}\ and\ \citenamefont {Woody}(1996)}]{lynds96a}%
  \BibitemOpen
  \bibfield  {author} {\bibinfo {author} {\bibfnamefont {L.}~\bibnamefont
  {Lynds}}\ and\ \bibinfo {author} {\bibfnamefont {B.~A.}\ \bibnamefont
  {Woody}},\ }\bibfield  {title} {\bibinfo {title} {Sub‐{Doppler}
  spectroscopy of {$^{89}$Y}},\ }\href {https://doi.org/10.1063/1.361420}
  {\bibfield  {journal} {\bibinfo  {journal} {Journal of Applied Physics}\
  }\textbf {\bibinfo {volume} {79}},\ \bibinfo {pages} {565} (\bibinfo {year}
  {1996})}\BibitemShut {NoStop}%
\bibitem [{\citenamefont {Riyadh}\ \emph {et~al.}(2023)\citenamefont {Riyadh},
  \citenamefont {Telfah}, \citenamefont {Jones}, \citenamefont {Bersson},
  \citenamefont {Cheng}, \citenamefont {Hu}, \citenamefont {Foote},\ and\
  \citenamefont {Liu}}]{liu23a}%
  \BibitemOpen
  \bibfield  {author} {\bibinfo {author} {\bibfnamefont {S.~M.~S.}\
  \bibnamefont {Riyadh}}, \bibinfo {author} {\bibfnamefont {H.}~\bibnamefont
  {Telfah}}, \bibinfo {author} {\bibfnamefont {I.}~\bibnamefont {Jones}},
  \bibinfo {author} {\bibfnamefont {J.}~\bibnamefont {Bersson}}, \bibinfo
  {author} {\bibfnamefont {C.}~\bibnamefont {Cheng}}, \bibinfo {author}
  {\bibfnamefont {S.-M.}\ \bibnamefont {Hu}}, \bibinfo {author} {\bibfnamefont
  {D.}~\bibnamefont {Foote}},\ and\ \bibinfo {author} {\bibfnamefont
  {J.}~\bibnamefont {Liu}},\ }\bibfield  {title} {\bibinfo {title}
  {Mid-{Infrared} {Doppler}-free saturation absorption spectroscopy of the {Q}
  branch of {CH$_4$} $\nu_3$=1 band using a rapid-scanning continuous-wave
  optical parametric oscillator},\ }\href
  {https://doi.org/10.1364/opticaopen.22138301.v1} {\bibfield  {journal}
  {\bibinfo  {journal} {Optica Open}\ ,\ \bibinfo {pages} {22138301}} (\bibinfo
  {year} {2023})}\BibitemShut {NoStop}%
\bibitem [{\citenamefont {Diouf}\ \emph {et~al.}(2022)\citenamefont {Diouf},
  \citenamefont {Tóbi\'as}, \citenamefont {van~der Schaaf}, \citenamefont
  {Cozijn}, \citenamefont {Salumbides}, \citenamefont {Cs\'asz\'ar},\ and\
  \citenamefont {Ubachs}}]{diouf22a}%
  \BibitemOpen
  \bibfield  {author} {\bibinfo {author} {\bibfnamefont {M.~L.}\ \bibnamefont
  {Diouf}}, \bibinfo {author} {\bibfnamefont {R.}~\bibnamefont {Tóbi\'as}},
  \bibinfo {author} {\bibfnamefont {T.~S.}\ \bibnamefont {van~der Schaaf}},
  \bibinfo {author} {\bibfnamefont {F.~M.~J.}\ \bibnamefont {Cozijn}}, \bibinfo
  {author} {\bibfnamefont {E.~J.}\ \bibnamefont {Salumbides}}, \bibinfo
  {author} {\bibfnamefont {A.~G.}\ \bibnamefont {Cs\'asz\'ar}},\ and\ \bibinfo
  {author} {\bibfnamefont {W.}~\bibnamefont {Ubachs}},\ }\bibfield  {title}
  {\bibinfo {title} {Ultraprecise relative energies in the (2 0 0) vibrational
  band of {H$_2^{16}$O}},\ }\href
  {https://doi.org/10.1080/00268976.2022.2050430} {\bibfield  {journal}
  {\bibinfo  {journal} {Molecular Physics}\ }\textbf {\bibinfo {volume}
  {120}},\ \bibinfo {pages} {e2050430} (\bibinfo {year} {2022})}\BibitemShut
  {NoStop}%
\bibitem [{\citenamefont {Parthey}\ \emph {et~al.}(2011)\citenamefont
  {Parthey}, \citenamefont {Matveev}, \citenamefont {Alnis}, \citenamefont
  {Bernhardt}, \citenamefont {Beyer}, \citenamefont {Holzwarth}, \citenamefont
  {Maistrou}, \citenamefont {Pohl}, \citenamefont {Predehl}, \citenamefont
  {Udem}, \citenamefont {Wilken}, \citenamefont {Kolachevsky}, \citenamefont
  {Abgrall}, \citenamefont {Rovera}, \citenamefont {Salomon}, \citenamefont
  {Laurent},\ and\ \citenamefont {H{\"{a}}nsch}}]{parthey11a}%
  \BibitemOpen
  \bibfield  {author} {\bibinfo {author} {\bibfnamefont {C.~G.}\ \bibnamefont
  {Parthey}}, \bibinfo {author} {\bibfnamefont {A.}~\bibnamefont {Matveev}},
  \bibinfo {author} {\bibfnamefont {J.}~\bibnamefont {Alnis}}, \bibinfo
  {author} {\bibfnamefont {B.}~\bibnamefont {Bernhardt}}, \bibinfo {author}
  {\bibfnamefont {A.}~\bibnamefont {Beyer}}, \bibinfo {author} {\bibfnamefont
  {R.}~\bibnamefont {Holzwarth}}, \bibinfo {author} {\bibfnamefont
  {A.}~\bibnamefont {Maistrou}}, \bibinfo {author} {\bibfnamefont
  {R.}~\bibnamefont {Pohl}}, \bibinfo {author} {\bibfnamefont {K.}~\bibnamefont
  {Predehl}}, \bibinfo {author} {\bibfnamefont {T.}~\bibnamefont {Udem}},
  \bibinfo {author} {\bibfnamefont {T.}~\bibnamefont {Wilken}}, \bibinfo
  {author} {\bibfnamefont {N.}~\bibnamefont {Kolachevsky}}, \bibinfo {author}
  {\bibfnamefont {M.}~\bibnamefont {Abgrall}}, \bibinfo {author} {\bibfnamefont
  {D.}~\bibnamefont {Rovera}}, \bibinfo {author} {\bibfnamefont
  {C.}~\bibnamefont {Salomon}}, \bibinfo {author} {\bibfnamefont
  {P.}~\bibnamefont {Laurent}},\ and\ \bibinfo {author} {\bibfnamefont {T.~W.}\
  \bibnamefont {H{\"{a}}nsch}},\ }\bibfield  {title} {\bibinfo {title}
  {Improved measurement of the {hydrogen} {$1S-2S$} transition frequency},\
  }\href {https://doi.org/10.1103/PhysRevLett.107.203001} {\bibfield  {journal}
  {\bibinfo  {journal} {Phys. Rev. Lett.}\ }\textbf {\bibinfo {volume} {107}},\
  \bibinfo {pages} {203001} (\bibinfo {year} {2011})}\BibitemShut {NoStop}%
\bibitem [{\citenamefont {Matveev}\ \emph {et~al.}(2013)\citenamefont
  {Matveev}, \citenamefont {Parthey}, \citenamefont {Predehl}, \citenamefont
  {Alnis}, \citenamefont {Beyer}, \citenamefont {Holzwarth}, \citenamefont
  {Udem}, \citenamefont {Wilken}, \citenamefont {Kolachevsky}, \citenamefont
  {Abgrall}, \citenamefont {Rovera}, \citenamefont {Salomon}, \citenamefont
  {Laurent}, \citenamefont {Grosche}, \citenamefont {Terra}, \citenamefont
  {Legero}, \citenamefont {Schnatz}, \citenamefont {Weyers}, \citenamefont
  {Altschul},\ and\ \citenamefont {H{\"a}nsch}}]{matveev13a}%
  \BibitemOpen
  \bibfield  {author} {\bibinfo {author} {\bibfnamefont {A.}~\bibnamefont
  {Matveev}}, \bibinfo {author} {\bibfnamefont {C.~G.}\ \bibnamefont
  {Parthey}}, \bibinfo {author} {\bibfnamefont {K.}~\bibnamefont {Predehl}},
  \bibinfo {author} {\bibfnamefont {J.}~\bibnamefont {Alnis}}, \bibinfo
  {author} {\bibfnamefont {A.}~\bibnamefont {Beyer}}, \bibinfo {author}
  {\bibfnamefont {R.}~\bibnamefont {Holzwarth}}, \bibinfo {author}
  {\bibfnamefont {T.}~\bibnamefont {Udem}}, \bibinfo {author} {\bibfnamefont
  {T.}~\bibnamefont {Wilken}}, \bibinfo {author} {\bibfnamefont
  {N.}~\bibnamefont {Kolachevsky}}, \bibinfo {author} {\bibfnamefont
  {M.}~\bibnamefont {Abgrall}}, \bibinfo {author} {\bibfnamefont
  {D.}~\bibnamefont {Rovera}}, \bibinfo {author} {\bibfnamefont
  {C.}~\bibnamefont {Salomon}}, \bibinfo {author} {\bibfnamefont
  {P.}~\bibnamefont {Laurent}}, \bibinfo {author} {\bibfnamefont
  {G.}~\bibnamefont {Grosche}}, \bibinfo {author} {\bibfnamefont
  {O.}~\bibnamefont {Terra}}, \bibinfo {author} {\bibfnamefont
  {T.}~\bibnamefont {Legero}}, \bibinfo {author} {\bibfnamefont
  {H.}~\bibnamefont {Schnatz}}, \bibinfo {author} {\bibfnamefont
  {S.}~\bibnamefont {Weyers}}, \bibinfo {author} {\bibfnamefont
  {B.}~\bibnamefont {Altschul}},\ and\ \bibinfo {author} {\bibfnamefont
  {T.~W.}\ \bibnamefont {H{\"a}nsch}},\ }\bibfield  {title} {\bibinfo {title}
  {Precision measurement of the {hydrogen} {$1S-2S$} frequency via a 920-km
  fiber link},\ }\href {https://doi.org/10.1103/PhysRevLett.110.230801}
  {\bibfield  {journal} {\bibinfo  {journal} {Phys. Rev. Lett.}\ }\textbf
  {\bibinfo {volume} {110}},\ \bibinfo {pages} {230801} (\bibinfo {year}
  {2013})}\BibitemShut {NoStop}%
\bibitem [{\citenamefont {Altmann}\ \emph {et~al.}(2018)\citenamefont
  {Altmann}, \citenamefont {Dreissen}, \citenamefont {Salumbides},
  \citenamefont {Ubachs},\ and\ \citenamefont {Eikema}}]{altmann18a}%
  \BibitemOpen
  \bibfield  {author} {\bibinfo {author} {\bibfnamefont {R.~K.}\ \bibnamefont
  {Altmann}}, \bibinfo {author} {\bibfnamefont {L.~S.}\ \bibnamefont
  {Dreissen}}, \bibinfo {author} {\bibfnamefont {E.~J.}\ \bibnamefont
  {Salumbides}}, \bibinfo {author} {\bibfnamefont {W.}~\bibnamefont {Ubachs}},\
  and\ \bibinfo {author} {\bibfnamefont {K.~S.~E.}\ \bibnamefont {Eikema}},\
  }\bibfield  {title} {\bibinfo {title} {Deep-ultraviolet frequency metrology
  of {H$_2$} for tests of molecular quantum theory},\ }\href
  {https://doi.org/10.1103/PhysRevLett.120.043204} {\bibfield  {journal}
  {\bibinfo  {journal} {Phys. Rev. Lett.}\ }\textbf {\bibinfo {volume} {120}},\
  \bibinfo {pages} {043204} (\bibinfo {year} {2018})}\BibitemShut {NoStop}%
\bibitem [{\citenamefont {Meek}\ \emph {et~al.}(2018)\citenamefont {Meek},
  \citenamefont {Hipke}, \citenamefont {Guelachvili}, \citenamefont
  {H\"{a}nsch},\ and\ \citenamefont {Picqu\'{e}}}]{meek18a}%
  \BibitemOpen
  \bibfield  {author} {\bibinfo {author} {\bibfnamefont {S.~A.}\ \bibnamefont
  {Meek}}, \bibinfo {author} {\bibfnamefont {A.}~\bibnamefont {Hipke}},
  \bibinfo {author} {\bibfnamefont {G.}~\bibnamefont {Guelachvili}}, \bibinfo
  {author} {\bibfnamefont {T.~W.}\ \bibnamefont {H\"{a}nsch}},\ and\ \bibinfo
  {author} {\bibfnamefont {N.}~\bibnamefont {Picqu\'{e}}},\ }\bibfield  {title}
  {\bibinfo {title} {Doppler-free {Fourier} transform spectroscopy},\ }\href
  {https://doi.org/10.1364/OL.43.000162} {\bibfield  {journal} {\bibinfo
  {journal} {Opt. Lett.}\ }\textbf {\bibinfo {volume} {43}},\ \bibinfo {pages}
  {162} (\bibinfo {year} {2018})}\BibitemShut {NoStop}%
\bibitem [{\citenamefont {Chandler}\ and\ \citenamefont
  {Houston}(1987)}]{chandler87a}%
  \BibitemOpen
  \bibfield  {author} {\bibinfo {author} {\bibfnamefont {D.~W.}\ \bibnamefont
  {Chandler}}\ and\ \bibinfo {author} {\bibfnamefont {P.~L.}\ \bibnamefont
  {Houston}},\ }\bibfield  {title} {\bibinfo {title} {Two‐dimensional imaging
  of state‐selected photodissociation products detected by multiphoton
  ionization},\ }\href {https://doi.org/10.1063/1.453276} {\bibfield  {journal}
  {\bibinfo  {journal} {The Journal of Chemical Physics}\ }\textbf {\bibinfo
  {volume} {87}},\ \bibinfo {pages} {1445} (\bibinfo {year}
  {1987})}\BibitemShut {NoStop}%
\bibitem [{\citenamefont {Clausen}\ \emph {et~al.}(2021)\citenamefont
  {Clausen}, \citenamefont {Jansen}, \citenamefont {Scheidegger}, \citenamefont
  {Agner}, \citenamefont {Schmutz},\ and\ \citenamefont {Merkt}}]{clausen21a}%
  \BibitemOpen
  \bibfield  {author} {\bibinfo {author} {\bibfnamefont {G.}~\bibnamefont
  {Clausen}}, \bibinfo {author} {\bibfnamefont {P.}~\bibnamefont {Jansen}},
  \bibinfo {author} {\bibfnamefont {S.}~\bibnamefont {Scheidegger}}, \bibinfo
  {author} {\bibfnamefont {J.~A.}\ \bibnamefont {Agner}}, \bibinfo {author}
  {\bibfnamefont {H.}~\bibnamefont {Schmutz}},\ and\ \bibinfo {author}
  {\bibfnamefont {F.}~\bibnamefont {Merkt}},\ }\bibfield  {title} {\bibinfo
  {title} {Ionization energy of the metastable 2 {$^1$S$_0$} state of {$^4$He}
  from {R}ydberg-series extrapolation},\ }\href
  {https://doi.org/10.1103/PhysRevLett.127.093001} {\bibfield  {journal}
  {\bibinfo  {journal} {Phys. Rev. Lett.}\ }\textbf {\bibinfo {volume} {127}},\
  \bibinfo {pages} {093001} (\bibinfo {year} {2021})}\BibitemShut {NoStop}%
\bibitem [{\citenamefont {Husmann}\ \emph {et~al.}(2021)\citenamefont
  {Husmann}, \citenamefont {Bernier}, \citenamefont {Bertrand}, \citenamefont
  {Calonico}, \citenamefont {Chaloulos}, \citenamefont {Clausen}, \citenamefont
  {Clivati}, \citenamefont {Faist}, \citenamefont {Heiri}, \citenamefont
  {Hollenstein}, \citenamefont {Johnson}, \citenamefont {Mauchle},
  \citenamefont {Meir}, \citenamefont {Merkt}, \citenamefont {Mura},
  \citenamefont {Scalari}, \citenamefont {Scheidegger}, \citenamefont
  {Schmutz}, \citenamefont {Sinhal}, \citenamefont {Willitsch},\ and\
  \citenamefont {Morel}}]{husmann21a}%
  \BibitemOpen
  \bibfield  {author} {\bibinfo {author} {\bibfnamefont {D.}~\bibnamefont
  {Husmann}}, \bibinfo {author} {\bibfnamefont {L.-G.}\ \bibnamefont
  {Bernier}}, \bibinfo {author} {\bibfnamefont {M.}~\bibnamefont {Bertrand}},
  \bibinfo {author} {\bibfnamefont {D.}~\bibnamefont {Calonico}}, \bibinfo
  {author} {\bibfnamefont {K.}~\bibnamefont {Chaloulos}}, \bibinfo {author}
  {\bibfnamefont {G.}~\bibnamefont {Clausen}}, \bibinfo {author} {\bibfnamefont
  {C.}~\bibnamefont {Clivati}}, \bibinfo {author} {\bibfnamefont
  {J.}~\bibnamefont {Faist}}, \bibinfo {author} {\bibfnamefont
  {E.}~\bibnamefont {Heiri}}, \bibinfo {author} {\bibfnamefont
  {U.}~\bibnamefont {Hollenstein}}, \bibinfo {author} {\bibfnamefont
  {A.}~\bibnamefont {Johnson}}, \bibinfo {author} {\bibfnamefont
  {F.}~\bibnamefont {Mauchle}}, \bibinfo {author} {\bibfnamefont
  {Z.}~\bibnamefont {Meir}}, \bibinfo {author} {\bibfnamefont {F.}~\bibnamefont
  {Merkt}}, \bibinfo {author} {\bibfnamefont {A.}~\bibnamefont {Mura}},
  \bibinfo {author} {\bibfnamefont {G.}~\bibnamefont {Scalari}}, \bibinfo
  {author} {\bibfnamefont {S.}~\bibnamefont {Scheidegger}}, \bibinfo {author}
  {\bibfnamefont {H.}~\bibnamefont {Schmutz}}, \bibinfo {author} {\bibfnamefont
  {M.}~\bibnamefont {Sinhal}}, \bibinfo {author} {\bibfnamefont
  {S.}~\bibnamefont {Willitsch}},\ and\ \bibinfo {author} {\bibfnamefont
  {J.}~\bibnamefont {Morel}},\ }\bibfield  {title} {\bibinfo {title}
  {{SI}-traceable frequency dissemination at 1572.06\,{nm} in a stabilized
  fiber network with ring topology},\ }\href
  {https://doi.org/10.1364/OE.427921} {\bibfield  {journal} {\bibinfo
  {journal} {Opt. Express}\ }\textbf {\bibinfo {volume} {29}},\ \bibinfo
  {pages} {24592} (\bibinfo {year} {2021})}\BibitemShut {NoStop}%
\bibitem [{\citenamefont {Drake}(1999)}]{drake99a}%
  \BibitemOpen
  \bibfield  {author} {\bibinfo {author} {\bibfnamefont {G.~W.~F.}\
  \bibnamefont {Drake}},\ }\bibfield  {title} {\bibinfo {title} {High precision
  theory of atomic helium},\ }\href
  {https://doi.org/10.1238/Physica.Topical.083a00083} {\bibfield  {journal}
  {\bibinfo  {journal} {Physica Scripta}\ }\textbf {\bibinfo {volume} {T83}},\
  \bibinfo {pages} {83} (\bibinfo {year} {1999})}\BibitemShut {NoStop}%
\bibitem [{\citenamefont {Patk{\'{o}}{\v{s}}}\ \emph
  {et~al.}(2021)\citenamefont {Patk{\'{o}}{\v{s}}}, \citenamefont {Yerokhin},\
  and\ \citenamefont {Pachucki}}]{patkos21a}%
  \BibitemOpen
  \bibfield  {author} {\bibinfo {author} {\bibfnamefont {V.}~\bibnamefont
  {Patk{\'{o}}{\v{s}}}}, \bibinfo {author} {\bibfnamefont {V.~A.}\ \bibnamefont
  {Yerokhin}},\ and\ \bibinfo {author} {\bibfnamefont {K.}~\bibnamefont
  {Pachucki}},\ }\bibfield  {title} {\bibinfo {title} {Complete {$\alpha^7m$}
  {L}amb shift of helium triplet states},\ }\href
  {https://doi.org/10.1103/PhysRevA.103.042809} {\bibfield  {journal} {\bibinfo
   {journal} {Phys. Rev. A}\ }\textbf {\bibinfo {volume} {103}},\ \bibinfo
  {pages} {042809} (\bibinfo {year} {2021})}\BibitemShut {NoStop}%
\bibitem [{\citenamefont {Wen}\ \emph {et~al.}(2023)\citenamefont {Wen},
  \citenamefont {Tang}, \citenamefont {Dong}, \citenamefont {Du}, \citenamefont
  {Hu},\ and\ \citenamefont {Sun}}]{wen23b}%
  \BibitemOpen
  \bibfield  {author} {\bibinfo {author} {\bibfnamefont {J.-L.}\ \bibnamefont
  {Wen}}, \bibinfo {author} {\bibfnamefont {J.-D.}\ \bibnamefont {Tang}},
  \bibinfo {author} {\bibfnamefont {J.-F.}\ \bibnamefont {Dong}}, \bibinfo
  {author} {\bibfnamefont {X.-J.}\ \bibnamefont {Du}}, \bibinfo {author}
  {\bibfnamefont {S.-M.}\ \bibnamefont {Hu}},\ and\ \bibinfo {author}
  {\bibfnamefont {Y.~R.}\ \bibnamefont {Sun}},\ }\bibfield  {title} {\bibinfo
  {title} {{Doppler}-free spectroscopy of an atomic beam probed in
  traveling-wave fields}} (\bibinfo {year} {2023}),\ \bibinfo {note}
  {{Abstract, International Conference on Precision Physics and Fundamental
  Physical Constants FFK. We thank Prof. Shui-Ming Hu, University of Science
  and Technology of China, Hefei, for drawing our attention to this systematic
  effect.}}\BibitemShut {Stop}%
\bibitem [{\citenamefont {Ritz}(1903)}]{ritz03a}%
  \BibitemOpen
  \bibfield  {author} {\bibinfo {author} {\bibfnamefont {W.}~\bibnamefont
  {Ritz}},\ }\bibfield  {title} {\bibinfo {title} {{Zur Theorie der
  Serienspektren}},\ }\href {https://doi.org/10.1002/andp.19033171003}
  {\bibfield  {journal} {\bibinfo  {journal} {Ann. Phys.}\ }\textbf {\bibinfo
  {volume} {317}},\ \bibinfo {pages} {264} (\bibinfo {year}
  {1903})}\BibitemShut {NoStop}%
\bibitem [{\citenamefont {Tiesinga}\ \emph {et~al.}(2021)\citenamefont
  {Tiesinga}, \citenamefont {Mohr}, \citenamefont {Newell},\ and\ \citenamefont
  {Taylor}}]{codata2018a}%
  \BibitemOpen
  \bibfield  {author} {\bibinfo {author} {\bibfnamefont {E.}~\bibnamefont
  {Tiesinga}}, \bibinfo {author} {\bibfnamefont {P.~J.}\ \bibnamefont {Mohr}},
  \bibinfo {author} {\bibfnamefont {D.~B.}\ \bibnamefont {Newell}},\ and\
  \bibinfo {author} {\bibfnamefont {B.~N.}\ \bibnamefont {Taylor}},\ }\bibfield
   {title} {\bibinfo {title} {{CODATA} recommended values of the fundamental
  physical constants: 2018},\ }\href
  {https://doi.org/10.1103/RevModPhys.93.025010} {\bibfield  {journal}
  {\bibinfo  {journal} {Rev. Mod. Phys.}\ }\textbf {\bibinfo {volume} {93}},\
  \bibinfo {pages} {025010} (\bibinfo {year} {2021})}\BibitemShut {NoStop}%
\bibitem [{\citenamefont {Drake}\ and\ \citenamefont
  {Swainson}(1991)}]{drake91a}%
  \BibitemOpen
  \bibfield  {author} {\bibinfo {author} {\bibfnamefont {G.~W.~F.}\
  \bibnamefont {Drake}}\ and\ \bibinfo {author} {\bibfnamefont {R.~A.}\
  \bibnamefont {Swainson}},\ }\bibfield  {title} {\bibinfo {title} {Quantum
  defects and the {$1/n$} dependence of {R}ydberg energies: {S}econd-order
  polarization effects},\ }\href {https://doi.org/10.1103/PhysRevA.44.5448}
  {\bibfield  {journal} {\bibinfo  {journal} {Phys. Rev. A}\ }\textbf {\bibinfo
  {volume} {44}},\ \bibinfo {pages} {5448} (\bibinfo {year}
  {1991})}\BibitemShut {NoStop}%
\bibitem [{\citenamefont {Rengelink}\ \emph {et~al.}(2018)\citenamefont
  {Rengelink}, \citenamefont {van~der Werf}, \citenamefont {Notermans},
  \citenamefont {Jannin}, \citenamefont {Eikema}, \citenamefont {Hoogerland},\
  and\ \citenamefont {Vassen}}]{rengelink18a}%
  \BibitemOpen
  \bibfield  {author} {\bibinfo {author} {\bibfnamefont {R.~J.}\ \bibnamefont
  {Rengelink}}, \bibinfo {author} {\bibfnamefont {Y.}~\bibnamefont {van~der
  Werf}}, \bibinfo {author} {\bibfnamefont {R.~P. M. J.~W.}\ \bibnamefont
  {Notermans}}, \bibinfo {author} {\bibfnamefont {R.}~\bibnamefont {Jannin}},
  \bibinfo {author} {\bibfnamefont {K.~S.~E.}\ \bibnamefont {Eikema}}, \bibinfo
  {author} {\bibfnamefont {M.~D.}\ \bibnamefont {Hoogerland}},\ and\ \bibinfo
  {author} {\bibfnamefont {W.}~\bibnamefont {Vassen}},\ }\bibfield  {title}
  {\bibinfo {title} {Precision spectroscopy of helium in a magic wavelength
  optical dipole trap},\ }\href {https://doi.org/10.1038/s41567-018-0242-5}
  {\bibfield  {journal} {\bibinfo  {journal} {Nature Physics}\ }\textbf
  {\bibinfo {volume} {14}},\ \bibinfo {pages} {1132} (\bibinfo {year}
  {2018})}\BibitemShut {NoStop}%
\bibitem [{\citenamefont {Yerokhin}\ \emph {et~al.}(2022)\citenamefont
  {Yerokhin}, \citenamefont {Patk{\'{o}}{\v{s}}},\ and\ \citenamefont
  {Pachucki}}]{yerokhin22a}%
  \BibitemOpen
  \bibfield  {author} {\bibinfo {author} {\bibfnamefont {V.~A.}\ \bibnamefont
  {Yerokhin}}, \bibinfo {author} {\bibfnamefont {V.}~\bibnamefont
  {Patk{\'{o}}{\v{s}}}},\ and\ \bibinfo {author} {\bibfnamefont
  {K.}~\bibnamefont {Pachucki}},\ }\bibfield  {title} {\bibinfo {title}
  {Relativistic {Bethe} logarithm for triplet states of helium-like ions},\
  }\href {https://doi.org/10.1140/epjd/s10053-022-00474-8} {\bibfield
  {journal} {\bibinfo  {journal} {Eur. Phys. J. D}\ }\textbf {\bibinfo {volume}
  {76}},\ \bibinfo {pages} {142} (\bibinfo {year} {2022})}\BibitemShut
  {NoStop}%
\bibitem [{\citenamefont {Yerokhin}\ \emph {et~al.}(2023)\citenamefont
  {Yerokhin}, \citenamefont {Patk{\'{o}}{\v{s}}},\ and\ \citenamefont
  {Pachucki}}]{yerokhin23a}%
  \BibitemOpen
  \bibfield  {author} {\bibinfo {author} {\bibfnamefont {V.~A.}\ \bibnamefont
  {Yerokhin}}, \bibinfo {author} {\bibfnamefont {V.}~\bibnamefont
  {Patk{\'{o}}{\v{s}}}},\ and\ \bibinfo {author} {\bibfnamefont
  {K.}~\bibnamefont {Pachucki}},\ }\bibfield  {title} {\bibinfo {title} {{QED}
  $m\alpha^7$ effects for triplet states of heliumlike ions},\ }\href
  {https://doi.org/10.1103/PhysRevA.107.012810} {\bibfield  {journal} {\bibinfo
   {journal} {Phys. Rev. A}\ }\textbf {\bibinfo {volume} {107}},\ \bibinfo
  {pages} {012810} (\bibinfo {year} {2023})}\BibitemShut {NoStop}%
\bibitem [{\citenamefont {Ashfold}\ and\ \citenamefont
  {Western}(2017)}]{ashfold17a}%
  \BibitemOpen
  \bibfield  {author} {\bibinfo {author} {\bibfnamefont {M.~N.~R.}\
  \bibnamefont {Ashfold}}\ and\ \bibinfo {author} {\bibfnamefont {C.~M.}\
  \bibnamefont {Western}},\ }\bibfield  {title} {\bibinfo {title} {Multiphoton
  {Spectroscopy}, {Applications}},\ }in\ \href
  {https://doi.org/10.1016/B978-0-12-409547-2.05032-0} {\emph {\bibinfo
  {booktitle} {Encyclopedia of {Spectroscopy} and {Spectrometry} ({Third}
  {Edition})}}},\ \bibinfo {editor} {edited by\ \bibinfo {editor}
  {\bibfnamefont {J.~C.}\ \bibnamefont {Lindon}}, \bibinfo {editor}
  {\bibfnamefont {G.~E.}\ \bibnamefont {Tranter}},\ and\ \bibinfo {editor}
  {\bibfnamefont {D.~W.}\ \bibnamefont {Koppenaal}}}\ (\bibinfo  {publisher}
  {Academic Press},\ \bibinfo {address} {Oxford},\ \bibinfo {year} {2017})\
  pp.\ \bibinfo {pages} {954--961}\BibitemShut {NoStop}%
\bibitem [{\citenamefont {Zare}(2012)}]{zare11a}%
  \BibitemOpen
  \bibfield  {author} {\bibinfo {author} {\bibfnamefont {R.~N.}\ \bibnamefont
  {Zare}},\ }\bibfield  {title} {\bibinfo {title} {My life with {LIF}: A
  personal account of developing laser-induced fluorescence},\ }\href
  {https://doi.org/10.1146/annurev-anchem-062011-143148} {\bibfield  {journal}
  {\bibinfo  {journal} {Annual Review of Analytical Chemistry}\ }\textbf
  {\bibinfo {volume} {5}},\ \bibinfo {pages} {1} (\bibinfo {year}
  {2012})}\BibitemShut {NoStop}%
\bibitem [{\citenamefont {Eppink}\ and\ \citenamefont
  {Parker}(1997)}]{eppink97a}%
  \BibitemOpen
  \bibfield  {author} {\bibinfo {author} {\bibfnamefont {A.~T. J.~B.}\
  \bibnamefont {Eppink}}\ and\ \bibinfo {author} {\bibfnamefont {D.~H.}\
  \bibnamefont {Parker}},\ }\bibfield  {title} {\bibinfo {title} {Velocity map
  imaging of ions and electrons using electrostatic lenses: Application in
  photoelectron and photofragment ion imaging of molecular oxygen},\ }\href
  {https://doi.org/10.1063/1.1148310} {\bibfield  {journal} {\bibinfo
  {journal} {Rev. Sci. Instr.}\ }\textbf {\bibinfo {volume} {68}},\ \bibinfo
  {pages} {3477} (\bibinfo {year} {1997})}\BibitemShut {NoStop}%
\end{thebibliography}

\end{document}